# Open-source code for manifold-based 3D rotation recovery of X-ray scattering patterns

Aliakbar Jafarpour

*Dept. of Biomolecular Mechanisms, Max-Planck Inst. for Medical Research, Jahnstr 29, 69120 Hiedelberg, Germany*
*jafarpour.a.j@ieee.org*

**Abstract:** Single particle 3D imaging with ultrashort X-ray laser pulses is based on collecting and combining the information content of 2D scattering patterns of an object at different orientations. Typical sample-delivery schemes leave little or no room for controlling the orientations. As such, the orientation associated with a given snapshot should be estimated after the experiment. Here we present an open-source code for the most rigorous technique having been reported in this context. Some practical issues along with proposed solutions are also discussed.

## 1. Introduction

*1.1. Context of the problem*

3D volumetric imaging of single biomolecules with X-ray lasers is an emerging field, listed as one of the 10 breakthroughs of the year by the Science Magazine [1]. It may be perceived as an imaging modality in-between the two established fields of X-ray crystallography and single particle imaging with electron microscopy. In the framework of the geometrical model of X-ray scattering [2], a 2D X-ray scattering pattern is a (spherical) slice through the 3D Fourier transform of the sought 3D density profile [3,4]. By appropriately orienting and combining such 2D slices, the full 3D Fourier transform (and with phase retrieval, also the sought real-space 3D density) is expected to be found.

Retrieval of the unknown 3D orientation associated with a given snapshot is a challenging problem. Some proposed approaches for solving this problem have been briefly introduced, compared, and contrasted in Appendix 1. Here, we focus on the most rigorous formulation presented so far, referred to as the *Manifold* technique in this contribution.

*1.2. Significance of the reported implementation*

Some necessary conditions for a successful analysis of a dataset with an orientation-recovery program are 1) the *pragmatic* validity of the technique, 2) the applicability of the technique to a given object, 3) a correct implementation of the concept and associated algorithms in the program, 4) optimization of *general* parameters of the program, 5) optimization (fine-tuning) of *object-specific* parameters of the program, and 6) further *data-specific* tuning of parameters to account for distortions or higher-order effects in experimental data.

Aspects such as accelerated computation and the possibility of working with small datasets are also important (and in some cases determining) factors, even though theoretically they do not affect the validity of the program. Of special significance is the sensitivity of the Manifold-based technique to uniform sampling [5,6] of the rotation space and the need for a very large dataset in the case of random sampling.

The conception, development, and validation of the Manifold technique are credited to Giannakis et al. [7]. The code presented and discussed here facilitates other aforementioned aspects. Starting from a working example (Section 3.3), one can introduce changes to different parameters to perform sensitivity analysis, or to fine-tune some very few parameters to adapt the program to a given dataset. The size of the test dataset and hence the associated computation cost have also been significantly reduced. This situation is to be contrasted with one, in which the code and all its parameters are simultaneously changed in a nearly-blind and inconclusive optimization with a large dataset.

*1.3. Outline*

This report has been structured as follows. In Section 2, the Manifold technique is briefly reviewed. The implemented Matlab program and a working example are introduced in Section 3, and conclusions are made in Section 4. Appendices 1-4 include more technical details and the complete Matlab script for an analysis.

**2. Manifold-based approach to 3D orientation recovery**

*2.1. Analytical formulation (forward problem)*

3D rotations of a given object change its scattering patterns. The combined information of all such patterns in the case of an *N*-pixel detector can be visualized as a hyper-surface in the *N*-dimensional pixel coordinate with 3 degrees of freedom. This hyper-surface includes information about 1) the object and 2) orientations. A suitable differential operator (measure of *distance*) can potentially disentangle these two sets of information and quantify relative rotations in terms of relative distances in a lower-dimensional space.

The simple yet significant conclusion of Ref. 7 is that such a dimensionality reduction and disentanglement is (under appropriate conditions, for example for objects that are not very anisotropic) possible. The resulting hyper-surface with *3 degrees of freedom* is indeed

*embedded in a 9D space*. The 9 coordinates of this space correspond (eventually) to the 9 elements of the rotation matrix associated with the sought orientation.

*2.2. Numerical solution (inverse problem)*

Starting with $N$ rotation-induced scattering patterns, one uses the following steps (as summarized graphically in Fig. 1):
- Forming the Distance Matrix $S_{N \times N}$, with each element $s_{i,j}$ being the difference (with a given measure) between two snapshots $i$ and $j$.
- Forming the *transition probability* matrix $P_{N \times N}$ from $S_{N \times N}$ (with *spectral graph analysis*, as detailed in *Tables 2 and 3 in Ref. 4*).
- Calculating the first 10 eigenvectors of the matrix $P_{N \times N}$ and saving the last 9 (non-trivial) eigenvectors as $\psi_{9 \times N}$
- Using nonlinear regression to find a matrix $c_{9 \times 9}$ correlating all eigenvectors $\psi_{(2:10),i}$ to the 9 elements of the sought rotation matrices $R_i$. The regression tries to enforce $9N$ linear equations $R_{9 \times N} = c_{9 \times 9} \psi_{9 \times N}$ and $2N$ nonlinear constraints (orthogonal and unity-determinant rotation matrices $r_{3 \times 3}^i = V2M\{R_{(1:9),i}\}$, where $V2M$ represents conversion of a 9-element vector to a 3×3 matrix).

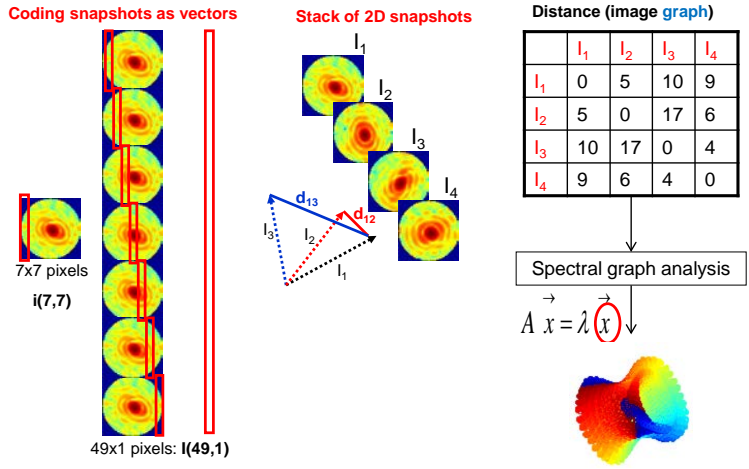

**Fig. 1**: Graphical visualization of the steps involved in the Manifold-based orientation recovery of X-ray scattering patterns of a single particle. The cylinder-looking manifold is the visualization of $(\psi_2, \psi_3, \psi_4)$ components of the Diffusion Coordinate, and the smoothly-varying color-code corresponds to known orientations.

## 3. The code and a working example

An important question is whether accumulated numerical errors in the estimation of the matrix $c_{9 \times 9}$ are caused by an inefficient optimization (nonlinear regression) or by an inefficient estimation of the Diffusion Coordinate (input data used by nonlinear regression). With simulated data, these two factors can be easily disentangled, as the $c_{9 \times 9}$ matrix can be estimated using *known* rotations. This makes it possible to optimize two major parameters (Sec. 3.2) without uncertainties about the success of a nonlinear optimization. As such, the program provides the possibility of estimation of $c_{9 \times 9}$ with and without *a-priori* rotation information.

*3.1. Overview of basic steps*
- Simulating snapshots corresponding to $N$ different orientations $\{I_{R_n}\}$

- Forming the *Distance Matrix* $S_{N \times N}$ from $N$ (simulated or measured) snapshots $\{I_{R_n}\}$
- Mapping the *Distance Matrix* $S_{N \times N}$ to a 9D (*Diffusion*) coordinate $\psi_{9 \times N}$, as described in *Tables 2 and 3 in Ref. 4.*
- Estimating the constant matrix $c_{9 \times 9}$, either with or without *a-priori* information
- Estimating the sought rotation matrices using the matrix $c_{9 \times 9}$
- Calculating a self-consistency measure of error (all mutual distances)

*3.2. Two major parameters: Size and (linear) calibration of a neighborhood*

For two images $x_i$ and $x_j$, a measure of similarity as a function of distance is defined as $W_{i,j} = W_{j,i} = \exp\left[-|x_i - x_j|^2/\epsilon\right]$. Furthermore, distances to the given image $x_i$, are only considered up to the $k$'th Nearest Neighbor ($kNN$). The main philosophy behind $KNN$ is to model the manifold around $x_i$ with as many "reliable" points as possible. Fortunately, it also results in significant reduction of computation costs. The pair $(\epsilon, k)$ has an important impact on appropriate conversion of a given set of images to a Similarity Matrix and subsequent analyses.

*3.3. Working example: Parameters*

An asymmetric object is simulated in real space within a 200×200×200 $nm^3$ cubic space, discretized with 31 points along each coordinate. Nonzero density values are limited to a maximum of 47% of the range along each coordinate. Each snapshot samples the diffraction volume with a 63×63 flat 2D grid. Grid sizes of the form $2^N - 1$ are used to have symmetric discrete grids including zero and also to use fast Fourier transform (fft) algorithms efficiently.

Geometrical and physical parameters of the simulated diffraction setup are chosen similar to typical soft X-ray experiments performed on biological samples; i.e., a 1024×1024 detector with a pixel pitch of $75\mu m$ placed at a distance of $z_D = 0.5m$ far from the interaction zone of the 3D object with X-ray photons with energy of $E = 620 eV$ ($\lambda = 2nm$).

Snapshots corresponding to $28^3 \sim 22,000$ rotations are generated by first rotating the object in the real space, and then sampling the corresponding diffraction volume with the camera. The rotations have a nearly-uniform deterministic distribution in SO(3); i.e., a Hopf fibration [8] with an 28×28×28 angular grid. In the program, the (Fourier-space) inversion symmetry may also be eliminated by synthesizing a (real-space) complex 3D density. However, in this test example, the complex factor is a constant.

*3.4. Working example: Results*

The program successfully calculates and uses the Diffusion coordinate. *Using the a-priori knowledge of rotation matrices*, it retrieves the orientations with a mean error of 5.9°, corresponding to all image pairs. The critical parameters of the diffusion map algorithm are $(\epsilon, k) = (0.7, 150)$. No manifold smoothing or any other processing of snapshots is employed. These results are obtained using known rotation matrices to estimate $c_{9 \times 9}$.

Using the normal procedure to extract both $R_{9 \times N}$ and $c_{9 \times 9}$ *with no a-priori information* increases the error from 5.9° to 6.4°. The increased error quantifies the negative impact of inefficient nonlinear regression (in a structured optimization landscape). There has been no special consideration for improving or optimizing the nonlinear regression (educated guess of initial value, multiple runs and evolutionary-like algorithms …), and there is room for improvement with such schemes.

The matrix $c_{9 \times 9}$ (within a scale factor) obtained after a nonlinear regression, along with the trend in the residual error as function of iteration are shown in Fig. 2. Of special interest is the capability of the nonlinear optimization in bypassing local optima (as for example after the fourth iteration), where a small slope of the "Residual vs. Iteration" curve is followed by an (abrupt) negative slope.

If the initial guess for $c_{9\times 9}$ is chosen as a random array with different seeds, the curves will look different (different trajectories in the optimization landscape will be traversed). The final estimated value for $c_{9\times 9}$, however, may or may not be precise.

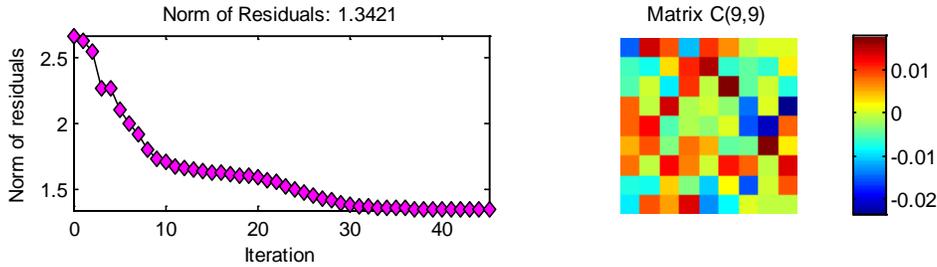

**Fig. 2:** (Left) The progress of the nonlinear regression as Residual error vs. iteration, and (right) the final result of nonlinear regression; i.e., the sought matrix $c_{9\times 9}$.

*3.5. Working example: Sensitivity of some parameters*

Decreasing the number of rotations by going from a 28×28×28 Hopf grid to a 24×24×24 one (from ~ 22,000 to ~ 14,000 snapshots) increases the error from 5.9° to 30°. This error has been obtained using *a-priori* rotation information and excludes additional contributions from nonlinear regression. These observations demonstrate (and quantify) the significant negative impact of sparse sampling of the rotation space on the performance of the algorithm. Since the error is defined as an inverse cosine function with extreme sensitivity at zero [7], significant changes of the argument (quaternion error) can result in only a small angular decrease.

Starting with the 22000-snapshot dataset, one can do a detailed sensitivity analysis by introducing parameterized randomness in orientations, noise in snapshots, number/position of missing pixels [10] and etc. to quantify the impact of each factor on the error. Note that by using a nearly-uniform *deterministic* distribution of rotations, we have decreased the minimum number of snapshots (compared to the case of random distribution [7]) by almost two orders of magnitude, while increasing the resolution.

By changing the $(\epsilon,k)$ pair and monitoring the mean error of estimations, the optimization landscapes is found to be highly structured. As such, a decreasing trend (especially corresponding to a subset of parameters) does not necessarily imply approaching the optimal point and should be interpreted with care.

The commutation of rotation and 3D Fourier transform is correct for a continuous problem. In a discrete problem with different resolutions in real- and Fourier spaces, the accumulated numerical errors of interpolation (associated with rotation in a grid) can be different. Simulating snapshots by rotating the camera and sampling the 3D diffraction volume will be not only faster, but also potentially more accurate.

*3.6. Working example: Computational considerations*

A special high-level parallelization scheme (matrix-based and without forming intermediate 3D arrays) [9] has made the calculation of the Distance Matrix very efficient, simple, and fast. The three tasks of 1) calculation of the Distance Matrix, 2) sorting it and forming the kNN Distance Matrix, and finally 3) calculation of the Diffusion Coordinate take each only a few seconds on a personal computer with 24GB RAM and Intel W3530 CPU (4 cores at 2.8GHz). Iterative nonlinear regression using Matlab's own *lsqnonlin* command (*Optimization* toolbox) is *inherently* parallelized (parallel computation on a single mluti-core machine, even without *Parallel Computing* or *Distributed Computing* toolboxes). It takes almost 1 second per iteration on a 32-core machine and less than 20 seconds per iteration on the aforementioned 4-core machine.

## 4. Concluding remarks

The Manifold-based technique for 3D orientation recovery of X-ray scattering patterns has not-only a rigorous theoretical foundation, but also a systematic procedure for the inverse problem. Despite the need and enthusiasm for applying this technique to real-life data, its widespread application seems to have been hindered (even with simulated data) by uncertainties about valid coding, parameterization, and applicability to a given object. An open-source code along with a working example provides a reliable point for sensitivity analysis and adapting the parameters to specific objects. The significant reduction of computation cost with nearly-uniform deterministic sampling of 3D rotations (for simulated data) can also be helpful for validation and optimization of programs implementing other (validated) orientation-recovery schemes.

**Appendix 1: Major aspects and comparative view of orientation-recovery techniques**

Figure 3 shows the major aspects of three orientation-recovery algorithms (Manifold [7], Common Arc [11,12], and GIPRAL [13]) used for the analysis of X-ray single particle scattering patterns.

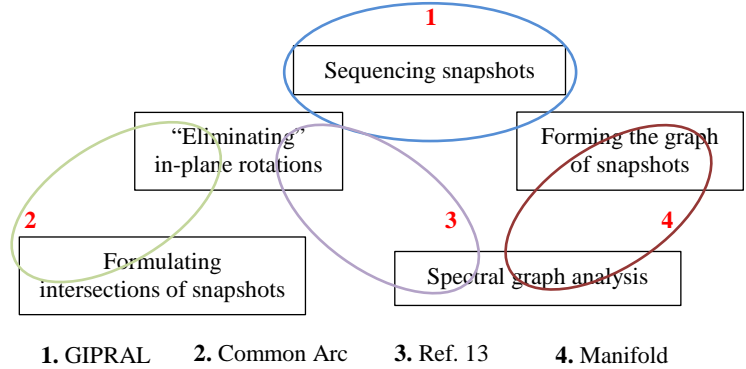

**1.** GIPRAL   **2.** Common Arc   **3.** Ref. 13   **4.** Manifold

**Figure 3**: Venn diagram showing the main aspects of four orientation-recovery algorithms.

The Manifold and the Common Arc techniques are different in many fundamental aspects. Also, datasets with nearly opposite properties (one including a few low-noise snapshots and the other with a large number of possibly-noisy snapshots) can be analyzed with one, but not the other. GIPRAL, however, has similarities to both techniques. These three techniques can be compared and contrasted, as detailed below and summarized in Table 1.

**Table 1: Comparison and contrast of three orientation-recovery algorithms**

|  | Number of images | Image Resolution | Based on image manifold | Based on image graph | Computation cost |
|---|---|---|---|---|---|
| Manifold | A lot | Low | Yes | Yes | High |
| GIPRAL | A lot | Moderate / High | No | Yes | Very high |
| Common Arc | ≥ 2 | Moderate / High | No | No | Low |

*A1.1. GIPRAL vs. Common Arc*

GIPRAL is similar to the Common Arc technique in the sense of 1) generating trivial rotations, 2) decomposing rotations into two such trivial rotations and one around an axis in the plane of camera (ZXZ or ZYZ Euler angle model), and 3) calculating correlations between all image pairs. GIPRAL is different from the Common Arc technique in the sense of 1) calculating correlations only for nearly-similar snapshots and using all pixels (as opposed to all snapshots and using the common arc pixels), 2) using the entire dataset at once with high-cost, yet parallelizable computation (as opposed to finding pairwise relative rotations separately, and subsequent clustering/filtering), and 3) not using any analytical formula (as opposed to analytically formulating the common arc).

*A1.2. GIPRAL vs. Manifold*

GIPRAL is similar to the Manifold technique in the sense of using a graph-based approach and evaluating pairwise distances for nearly similar snapshots. GIPRAL is different from the Manifold technique in the sense of 1) not *using* (and only *interpreting*) the image graph as a mesh on the image manifold, 2) using an inherently discrete approach, Dijkstra's algorithm, to define and quantify large-scale distances on the graph with no assumption about the manifold (as opposed to considering the graph as a dense-enough mesh on a smooth twice-differentiable 3-manifold of images with a given definition of metric), and 3) mapping the topological information of the reduced graph to geometrical 1D rotations with the assumptions of dense sampling and full angular coverage (as opposed to approximating a continuous metric on a 3-manifold using the entire graph).

*A1.3. Further overlaps and implication of the Hairy Ball Theorem*

For the sake of completeness (alternative combinations of the aspects shown in Figure 2), a fourth orientation-recovery technique [14], developed in the context of electron-microscopy 3D imaging, has also been specified. Of special significance is the implication of the Hairy Ball Theorem for 3D orientation recovery problems, both from a rigorous and a practical perspective [14]. This topic can be insightful for further developments of the techniques dealing with X-ray scattering patterns that explicitly model in-plane rotations [11,12,13].

## Appendix 2. Forming and processing the Distance Matrix

### A2.1. Spurious asymmetry induced by kNN

Using the common $k$-nearest-neaighbor ($kNN$) approach to limit distances to small values can cause an inconsistency in pairwise distances. The distance between a given image $x_i$ and its $k$ nearest neighbors are all finite numbers. However, these neighbors have $k$ closest neighbors of their own, potentially excluding the given image $x_i$. As such, the distance between these points and the given image can be set to infinity; i.e., $S_{i,j} = |x_i - x_j| < \infty$ but $S_{j,i} = \infty$. This *spurious* asymmetry (inconsistency) can cause serious numerical errors. One solution is to simply modify the final calculated similarity matrix $W = e^{-S^2/\epsilon}$ as $W \leftarrow (W + W')/2$ to enforce symmetry. While helpful, it still distorts the reliable information. An alternative approach is to put a threshold on the *value* of distance and not on the number of neighbors.

### A2.2. Distance histogram

The Manifold approach tries to approximate the real manifold with a smooth one. Smoothing a corrugated manifold is equivalent to discarding (or locally averaging) the regions with very small distances. The median of all values of "the first closest neighbor" is a measure of the average size of a cell on the discrete manifold. Snapshots generating distance values significantly smaller than this number should be either discarded, or locally averaged with their closest neighbors.

Distances considerably smaller than the average distance can be associated with the same orientation contaminated with two different noise components, or a sharp (non-smooth) corner on the manifold (which should be smoothed).

Local histogram-based filtering is an alternative for the "self-tuning" Kernel method [7] that sacrifices the image resolution everywhere on the manifold.

### A2.3. Round distance

Using the point-wise difference between images or the *Euclidean distance* is equivalent to approximating the image manifold locally by a plane. Modeling the local curvature requires a more tedious and seemingly impractical approach. This is indeed the task to be accomplished by spectral graph analysis; to use local similarities to estimate global features. However, in doing so, this task will be easier with better estimations of the local similarities, even with some phenomenological estimations of the local curvature.

One such approximation can be the assumption of isotropicity, similar to the rotation manifold. The rotation manifold is (the upper half of) the unit hypersphere in the 4D space (of quaternions) $q_x^2 + q_y^2 + q_z^2 + q_t^2 = 1$. Over this manifold (or similar round manifolds with other dimensions), the geodesic or *round distance* is $d_{round} = 2\sin^{-1}(d_{Eulcidean}/2)$. Note that $d_{Eulcidean}$, as defined here, is the Euclidean distance between two images after each being normalized to its norm.

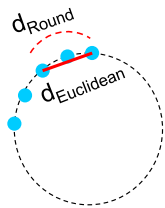

**Figure 4**: Euclidean and Round distances on the 1D rotation manifold, embedded in the 2D space. A similar relation holds in the case of 3D rotation manifold, embedded in the 4D space.

Alternative *phenomenological* models as a scalar functions $f(x)$ with the asymptotic behavior $f(x) \to x$ for small $x$ can also be used to define $d = f(d_{Euclidean})$. However, an educated guess or an intuitive justification may be less straightforward.

*A2.4. Offset rotation and degeneracy of optimization landscape*

The equation $R_{9 \times N} = c_{9 \times 9} \psi_{9 \times N}$ can be multiplied from the left side by any 9×9 non-singular matrix and modify both $c_{9 \times 9}$ and $R_{9 \times N}$. Geometrically, an offset 3D rotation can be added to all rotations, leaving relative orientations intact. It implies a degeneracy in the solution landscape and unnecessary local optima. An alternative optimization landscape may be used by explicit consideration of a reference orientation; i.e., forcing one of the rotation matrices to be $I_{3 \times 3}$. This snapshot (which is not unique) can be chosen using the sorted distance matrix, as one with a moderate (close to median) distance to its neighbors.

**Appendix 3: Uniform and enhanced samplings**

*A3.1. Nearly-uniform <u>deterministic</u> partitioning of the rotation space ($S^3$)*

Random sampling of the rotation-space (with uniform probability distribution function) is a meaningful and compromising solution to two different problems with different requirements. It is not necessarily a realistic model of practical "random" orientations (with preferred orientations due to anisotropicity). It is not the most suitable choice for validation of a code and optimization of parameters, either.

The core of the Manifold-based algorithm for orientation-recovery tries to use *discrete* data to approximate the main directions (eigenvectors) associated with a *continuous* operator (Laplace-Beltrami). However, with non-uniform sampling of discrete data (3D rotations), a different operator would be approximated. As such, an important pre-processing step of the complete algorithm is accounting for non-uniform sampling. It is done using the *anisotropic* Kernel matrix [6].

This trick (with anisotropic Kernel matrix parameter $\alpha = 1$) [5,6] is successful with a reasonably high density of sampled points even at the sparsest regions of the manifold. In practical terms, dense sampling means almost 2,000,000 simulated images with a low resolution of 40×40 [7] and considerably more images, when the resolution is increased.

This constraint (on the minimum number of required snapshots) can be significantly relaxed, when (nearly-) uniform *deterministic* sampling of 3D rotation is employed. Hopf fibration is a special mapping of the rotation space with such a nearly-uniform deternministic sampling [8]. When Hopf fibration and the common ZYZ Euler angle formulations are both expressed in terms of quaternions, it can be shown that for three angles $(\psi, \theta, \phi)$, the following relation holds:

$$Hopf(\psi, \theta, \phi) = Euler_{ZYZ}(\psi + \phi, \theta, -\phi)$$

In brief, starting from a uniform $(\psi, \theta, \phi)$ grid and then conversion to Euler angles using the above equation, one can have a nearly-uniform singularity-free partitioning of the rotation space with Euler angles formulation. Note that $\Delta\psi = \Delta\phi = 2\Delta\theta = 2\pi$.

*A3.2. Enhancing the sampling density of the rotation space*

A crucial assumption of the Manifold technique is that the spatial resolution in sampling the rotation space (number and distribution of snapshots) is high enough compared to the spatial resolution of snapshots. From another perspective, with a given set of rotations, one can (and in many cases should) down-sample snapshots to low resolutions.

One way to avoid excessive down-sampling of experimental data is to increase the effective density of snapshots sampled on the rotation manifold by generating in-plane-rotated versions of existing snapshots, as in the GIPRAL technique [13] (in combination with manifold-smoothing [7] or similar averaging schemes). Note that the already-high cost of computation of the Distance Matrix will become even higher with these additional snapshots.

**Appendix 4: Matlab implementation of the key formulae**

```matlab
%% Major functions
function []=Main(varargin)
    %Initialization
    close all force;
    clear all;
    clc;

    %Similarity Matrix parameters
    k=150;
    Epsilon=0.7;

    %Whether to use known rotation matrices to estimate c(9,9)
    aPrioriFlag=0;

    %Main menu
    switch menu('Please select the operation',...
            'Generation of snapshots',...
            'Calculation of Distance Matrix',...
            'Estimation of Diffusion Coordinate');
        case 1
            Rotated_Images();
        case 2
            Calculate_Distance_Matrix(k);
            disp(['k=' num2str(k)]);
        case 3
            Calculate_Diffusion(k,Epsilon,aPrioriFlag);
            disp(['k=' num2str(k) ', Epsilon=' num2str(Epsilon)]);
    end
end
function Calculate_Distance_Matrix(k)
    Wait_Bar=waitbar(0,'Loading snapshots');drawnow;
    A=importdata('./Images.mat');
    close (Wait_Bar);drawnow;
    N_orient=size(A,2); %A: Images
    A=A-repmat(mean(A,2),1,size(A,2));
    Wait_Bar=waitbar(0,'Calculating the {\bfRound} Distance Matrix');
    drawnow;
    A=A./repmat(sqrt(sum(A.^2,1)),size(A,1),1); %Unity-norm Images
    A=real(acos(A'*A)).^2;   %A <-- Round distance of Images ^2
    S2=randn(N_orient,k);   %Initializing kNN Distance ^2
    N=randn(N_orient,k);    %Initializing kNN Indices
    Wait_Bar=waitbar(0,Wait_Bar,['Sorting distances (kNN, k=' ...
        num2str(k) ')']);drawnow;
    for cntr=1:N_orient
        [YY,II]=sort(A(cntr,:),'ascend');
        N(cntr,:)=II(1:k);
        S2(cntr,:)=YY(1:k);
        if ~mod(cntr,50)
            waitbar(cntr/N_orient,Wait_Bar);drawnow;
        end
    end
    clear A;
    save S2.mat S2 -v7.3;
    save N.mat N -v7.3;
    close(Wait_Bar);drawnow;
end
function []=Calculate_Diffusion(k,Epsilon,aPrioriFlag)
    Wait_Bar=waitbar(0,'Loading matrices');drawnow;
    S2=importdata('./S2.mat');
    N=importdata('./N.mat');
    Rotation_Axis=importdata('./Axis.mat');
    Rotation_R=Axis2RotUnfold(Rotation_Axis);
    Wait_Bar=waitbar(0.5,Wait_Bar,...
        'Finding the Diffusion Coordinate \psi');drawnow;
    [Ps,Lambda]=Diffusion_Coordinate(S2,N,Epsilon);
```

```matlab
        close(Wait_Bar);drawnow;
        if isnan(Lambda)
            disp(['Problem in convergence of eigenvalues with Epsilon=' ...
                num2str(Epsilon) ' and k=' num2str(k)]);
        elseif ~isreal(Ps(:))
            disp(['Complex eigenvector found with Epsilon=' ...
                num2str(Epsilon) ' and k=' num2str(k)]);
        else
            Plot_Diffusion_Coordinate(Ps,Lambda,Rotation_Axis);
            save Ps.mat Ps -v7.3;
            save Lambda.mat Lambda -v7.3;
            [c,c0]=DM_Fit_All(Ps,Rotation_R,aPrioriFlag);
            save c0.mat c0 -v7.3;
            save c.mat c -v7.3;
        end
end
function [c,c0]=DM_Fit_All(Ps,Rotation_R,aPrioriFlag)
    if aPrioriFlag
        aPrioriText=' {\bfwith} ';
    else
        aPrioriText=' {\bfwithout} ';
    end
    WB=waitbar(0,['Finding C' aPrioriText 'known rotations']);drawnow;
    Ps=Ps(:,2:end);
    Rotation_R=Rotation_R';
    if aPrioriFlag
        c0=Ps\Rotation_R;
        c=c0;
    else
        c0_1D=rand(81,1);
        c_scale=5;
        LB=[];
        UB=[];
        Options=Opt_param(c_scale,c0_1D);
        [c_1D,ResNorm,Residual,ExitFlag]= ...
            lsqnonlin(@(C)OR_Func(C,Ps'),c0_1D*c_scale,LB,UB,Options);
        c=pinv(reshape(c_1D,[9 9]));
        c0=pinv(reshape(c0_1D,[9 9]));
    end
    figure;imagesc(c);axis equal;axis off;title('Matrix C(9,9)');colorbar;
    WB=waitbar(0,WB,['Assessment of estimation (' aPrioriText ...
        'known rotations)']);drawnow;
    Recon_R=Ps*c;
    N=size(Ps,1);
    for cntr=1:N
        Temp_00=Polar_Decompose(reshape(Recon_R(cntr,:),[3 3]));
        Recon_R(cntr,:)=Temp_00(:);
    end
    Q_1=rand(4,N);      %Initializing "Eestimated" Quaternions of rotations
    Q_2=rand(4,N);      %Initializing "Known" Quaternions of rotations
    for cntr=1:N
        r0=reshape(Recon_R(cntr,:),[3 3]);
        Q_1(1:4,cntr)=RotMat2Quat(r0);
        r0=reshape(Rotation_R(cntr,:),[3 3]);
        Q_2(1:4,cntr)=RotMat2Quat(r0);
        if ~mod(cntr,5)
            WB=waitbar(cntr/N,WB,'Forming rotation arrays');drawnow;
        end
    end
    WB=waitbar(0,WB,'Assessment of estimated {\bfQuaternions}');drawnow;
    TempA=real(acos(abs(Q_1'*Q_1)));    %Pairwise geodesic distances
    TempB=real(acos(abs(Q_2'*Q_2)));    %Pairwise geodesic distances
    Sigma_T=(180/pi)*2*sum(abs(TempA(:)-TempB(:)))/(N*(N-1));
    disp('Measure of error in Relative Orientations of All Pairs')
    disp(['Sigma_All_Pairs: ' num2str(Sigma_T) ' degrees'])
    close(WB);drawnow;
```

```matlab
        end

%% Calculating the diffusion coordinate
function W=S2_2_W_Matrix(S2,N,Epsilon)
    [N_orient,d]=size(S2);
    S2_Max=median(S2(:,5));

    Index=1;
    Dist_Max=min(S2(:,d));
    while max(S2(:,Index)) < Dist_Max
        Index=Index+1;
    end
    Dist_Thr=max(S2(:,Index-1));
    S2(S2 > Dist_Thr) = inf;

    Epsilon=Epsilon*S2_Max;
    S2=double(S2);
    N=double(N);
    W=sparse(repmat((1:N_orient)',1,d),N,exp(-S2/Epsilon));
    %W=(W+W')/2;
end
function [Ps,Lambda]=Diffusion_Coordinate(S2,N,Epsilon)
    [N_orient,~]=size(S2);
    [P_ep,D]=DM_Cov(AnIsoNorm(S2_2_W_Matrix(S2,N,Epsilon)));

    opts.disp=0;     %2
    opts.v0=1-5e-4*(0:(N_orient-1))';
    [Ps,Lambda,Eigen_Flag]=eigs(P_ep,10,'LM',opts);
    if ~Eigen_Flag
        [Lambda,Index]=sort(diag(Lambda),'descend');
        Ps=Ps(:,Index);
        Lambda=Lambda(Index);
    else
        disp('Error in eigenvalue calculation');
        Lambda=nan;
    end
    for cntr=1:size(Ps,2)
        Ps(:,cntr)=Ps(:,cntr)*Lambda(cntr);
    end
    Ps=D*Ps;
end
function W=AnIsoNorm(W)
    N_orient=size(W,1);
    Min=1/N_orient;
    Q_Alpha=sum(W,2);
    Q_Alpha(Q_Alpha < Min ) = Min;
    Q_Alpha=spdiags(1./Q_Alpha,0,N_orient,N_orient);
    W=Q_Alpha*W*Q_Alpha;
    W=(W+W')/2;
end
function [W,D]=DM_Cov(W)
    N_orient=size(W,1);
    Min=1/N_orient;
    D=sum(W,2);
    D(D < Min ) = Min;
    D=spdiags(1./sqrt(D),0,N_orient,N_orient);
    W=D*W*D;
    W=(W+W')/2;
end

%% Setting the nonlinear optimization parameters
function Options=Opt_param(varargin)
    c_scale=varargin{1};
    Options=optimset();
    Options.TolFun=1e-20*c_scale;
```

```matlab
        Options.DiffMaxChange=1e5*c_scale;
        Options.Display='final';
        Options.MaxFunEvals=1e8;
        Options.MaxIter=250;    %1000
        Options.PlotFcns=@optimplotresnorm;
        %Options.PlotFcns=@optimplotx;
        Options.Algorithm='trust-region-reflective';
%       Options.Algorithm='levenberg-marquardt';
    end

    %% Miscellaneous plots of the diffusion coordinate
    function Plot_Diffusion_Coordinate(Ps,Lambda,Rotation)
        Plot_Eigenvalue_Eigenvector(Ps,Lambda)
        Plot_Individual_EV(Ps,Lambda)
        Plot_Diffusion_Statistics(Ps)
        Plot_Corr_Ps(Ps);
        Plot_Psi234_ColorRot(Ps,Lambda,Rotation,'Axis')
    end
    function Plot_Eigenvalue_Eigenvector(Ps,Lambda)
        figure
        set(gca,'NextPlot','replacechildren');
        subplot(211)
        imagesc(Ps)
        colorbar
        title('Diffusion map eigenfunctions {\psi_i}')
        subplot(212)
        plot(Lambda,'-*')
        colorbar
        title('Diffusion map eigenvalues')
    end
    function Plot_Diffusion_Statistics(Ps)
        figure
        set(gca,'NextPlot','replacechildren');
        subplot(221)
        bar(Ps(:));
        title(['First 10 \psi, min=' num2str(min(Ps(:))) ...
            ', Max=' num2str(max(Ps(:)))]);
        PsNorm=Ps(:,1).^2;
        for cntr=2:10
            PsNorm=PsNorm+Ps(:,cntr).^2;
        end
        PsNorm=sqrt(PsNorm);
        subplot(222)
        bar(PsNorm);
        title(['10-element norm, min=' num2str(min(PsNorm)) ...
            ', Max=' num2str(max(PsNorm))]);
        subplot(223)
        histfit(PsNorm,500,'logistic');
        title('10-element norm histogram');
        subplot(224)
        Index=(1:size(PsNorm,1))';
        Index=2*pi*Index/max(Index);
        polar(Index,PsNorm,'b');
        title('10-element norm polar histogram');
        hold on
        polar(Index,mean(PsNorm)*ones(size(PsNorm)),'r--');
        hold off
    end
    function Plot_Individual_EV(Ps,Lambda)
        figure
        set(gca,'NextPlot','replacechildren');
        for cntr=1:9
            subplot(3,3,cntr)
            plot(Ps(:,cntr+1).*Lambda(cntr+1))
            title(['\psi_' num2str(cntr+1)])
        end
```

```matlab
        end
        function Plot_Psi234_ColorRot(Ps,Lambda,Rot,Text)
            for cntr=1:size(Rot,1)
                figure
                set(gca,'NextPlot','replacechildren');
                
                subplot(221)
                scatter(Ps(:,2).*Lambda(2),Ps(:,3).*Lambda(3),20,Rot(cntr,:))
                title(['Phase plane: \psi_2 vs. \psi_3 - Color-coded by ' ...
                    Text ' ' num2str(cntr)]);
                colorbar
                
                subplot(222)
                scatter(Ps(:,2).*Lambda(2),Ps(:,4).*Lambda(4),20,Rot(cntr,:))
                title(['Phase plane: \psi_2 vs. \psi_4 - Color-coded by ' ...
                    Text ' ' num2str(cntr)]);
                colorbar
                
                subplot(223)
                scatter(Ps(:,3).*Lambda(3),Ps(:,4).*Lambda(4),20,Rot(cntr,:))
                title(['Phase plane: \psi_3 vs. \psi_4 - Color-coded by ' ...
                    Text ' ' num2str(cntr)]);
                colorbar
                
                subplot(224)
                scatter3(Ps(:,2).*Lambda(2),Ps(:,3).*Lambda(3),...
                    Ps(:,4).*Lambda(4),20,Rot(cntr,:))
                title(['Phase plane: \psi_2-\psi_3-\psi_4 - Color-coded by ' ...
                    Text ' ' num2str(cntr)]);
                colorbar
            end
            drawnow;
        end
        function Plot_Corr_Ps(Ps)
            figure
            subplot(4,1,1)
            plot(real(xcov(Ps(:,2),Ps(:,2))))
            title('Covariance of \psi_2 and \psi_2');
            legend('Auto correlation')
            ylim([-0.2 1.1])
            subplot(4,1,2)
            plot(real(xcov(Ps(:,2),Ps(:,3))))
            title('Covariance of \psi_2 and \psi_3');
            legend('Cross correlation')
            ylim([-0.2 1.1])
            subplot(4,1,3)
            plot(real(xcov(Ps(:,2),Ps(:,4))))
            title('Covariance of \psi_2 and \psi_4');
            legend('Cross correlation')
            ylim([-0.2 1.1])
            subplot(4,1,4)
            plot(real(xcov(Ps(:,3),Ps(:,4))))
            title('Covariance of \psi_3 and \psi_4');
            legend('Cross correlation')
            ylim([-0.2 1.1])
            drawnow;
        end
        
        %% Imposing the rotation matrix constraints to find {c}
        function G_Functional=OR_Func(c,PsMod)
            %Note: PsMod=Ps(:,1+(1:NPsi))'
            r=size(PsMod,2);
            r_MP5=1/sqrt(r);
            N_c=9;
            c_Temp=reshape(c,[N_c N_c]);
```

```matlab
        Temp=zeros(N_c,N_c);      %Memory allocation
        Temp2=randn(N_c^2,1);     %Memory allocation
        I=eye(3);
        
        G_Functional=zeros(r,1);
        R_Big=c_Temp*PsMod;
        for cntr_l=1:r
            R=reshape(R_Big(:,cntr_l),[3 3]);
            Temp=R'*R-I;
            Temp2=Temp(:);
            G_Functional(cntr_l)=sqrt(Temp2'*Temp2)+abs(det(R)-1);%L2
            %G_Functional(cntr_l)=sum(abs(Temp(:)))+abs(det(R)-1);   %L1
        end
        G_Functional=sqrt(G_Functional)*r_MP5;
    end

%% Generating snapshots
function []= Rotated_Images()
    % Input parameters
    N_loop=28^3;     %Cube of an "even" integer
    Experiment=Experiment_Parameters();
    
    % Loading the object
    WaitBar=waitbar(0,'Generating the 3D object');drawnow;
    Protein=Load_Protein();
    
    % Separating Object values (at voxels) and space coordinates
    Grid_3D=Protein.Grid_3D;
    ED=Protein.ED;
    clear Protein;
    
    % Imaging w/ initial orientation
    N_p=Experiment.N_p;
    N_p2=N_p^2;
    
    % Loop
    WaitBar=waitbar(0,WaitBar,'Generating Rotation Matrices');drawnow;
    R=AllRotMatrices(N_loop);
    
    R_size=[3 3];
    WaitBar=waitbar(0,WaitBar,'Memory allocation');drawnow;
    Images=randn(N_p2,N_loop);   %Memory allocation
    WaitBar=waitbar(0,WaitBar,['Generating ' num2str(N_loop) ...
        ' snapshots']);drawnow;
    
    [Lambda,zD,Width,N]=Extract_ExpParam(Experiment);
    [Length,Number]=Extract_Coordinates(Grid_3D);
    
    [~,k]=FourierScaledAxes(Number,Length);
    
    % Camera coordinate (k-space)
    [Camera_x,Camera_y]=meshgrid((Width/(N-1))*((1:N)-(N+1)/2));
    Circle_Index=((Camera_x.^2+Camera_y.^2) > (Width/2)^2);
    Temp=Lambda*sqrt(Camera_x.^2+Camera_y.^2+zD^2);
    Q_x=Camera_x./Temp;
    Q_y=Camera_y./Temp;
    Q_z=(zD./Temp-1/Lambda);
    
    for cntr=1:N_loop
        ED_rot=RotateStructureIndex(ED,reshape(R(:,cntr),R_size));
        Camera_I=interp3(k.x,k.y,k.z,Shift_FFT(abs(fftn(ED_rot))),...
            Q_x,Q_y,Q_z,'linear',0);
        Camera_I(Circle_Index)=0;
        Images(:,cntr)=reshape(Camera_I,[N_p2 1]);
        if ~mod(cntr,50)
```

```matlab
            waitbar(cntr/N_loop,WaitBar,['Generating snapshot ' ...
                num2str(cntr) ' out of ' num2str(N_loop)]);drawnow;
        end
    end
    WaitBar=waitbar(0,WaitBar,'Saving snapshots');drawnow;
    save Images.mat Images  '-v7.3';
    close(WaitBar);drawnow;
end
function Experiment=Experiment_Parameters()
    N_P_NoBin=1024;
    Experiment=struct;
    Experiment.N_p=63; %number of pixels along each coordinate
    Experiment.Pixel=75e-6;
    Experiment.zD=0.5;    %0.738
    Experiment.Lambda=2*1e-9;    %Doubled! 2*1.03e-9
    Experiment.SuperPixel=Experiment.Pixel*(N_P_NoBin/Experiment.N_p);
    Experiment.Width=Experiment.SuperPixel*Experiment.N_p;
end
function [Lambda,zD,Width,N]=Extract_ExpParam(Experiment)
    Lambda=Experiment.Lambda;
    zD=Experiment.zD;
    Width=Experiment.Width;
    N=Experiment.N_p;
end

%% Synthesing a 3D object in real-space
function Protein=Load_Protein(varargin)
    Protein_Source=2;
    switch Protein_Source
        case 1
            Protein_File='Protein.mat';
            WaitBar=waitbar(0,'Loading protein data');
            pause(1e-3);
            Protein=load(Protein_File);
            Protein=Protein.Protein;
            delete(WaitBar);
            pause(1e-3);
        case 2
            if nargin==1
                Protein=IR_3D(varargin{1});
            elseif ~nargin
                Protein=IR_3D();
            end
    end
end
function Protein=IR_3D(varargin)
    if ~nargin
        close all;
        %clear all;
        clc;
        pause(1e-6);
    end
    N1=31;
    N2=N1;
    N3=N1;
    U=((1:N1)-(N1+1)/2)/(N1/2);
    V=((1:N2)-(N2+1)/2)/(N2/2);
    W=((1:N3)-(N3+1)/2)/(N3/2);
    [x,y,z]=meshgrid(U,V,W);
    A=x/0.47;
    B=y/0.37;
    C=z/0.29;
    if nargin==1
        Scale=varargin{1};
        A=A/Scale;
```

```matlab
            B=B/Scale;
            C=C/Scale;
        end
        F=(1-0.4*((A-0.15).^2+(B+0.2).^2+(C-0.1).*(A-0.15).*(B+0.2)));
        F(  (cos(20*pi*(x-z-0.2).*abs(y+z+0.1).*abs(z-0.3)) < 0.2) | ...
            (A.^2+B.^2+C.^2 >1)   | (F<0) )=0;
        if ~nargin
            Protein=struct;
            Protein.ED=F;
            clear F;
            Factor=2e-7; %3e-7
            Protein.Grid_3D.x=x*Factor;
            Protein.Grid_3D.y=y*Factor;
            Protein.Grid_3D.z=z*Factor;
        else
            Protein=F;
        end
end
function F=Shift_FFT(F)
    N=size(F);
    Nh=(N-1)/2;
    for cntr=1:3
        Index{cntr}=[(Nh(cntr)+1):N(cntr),1:Nh(cntr)];
    end
    F=F(Index{1},Index{2},Index{3});
end
function [Nyquist,k]=FourierScaledAxes(Number,Length)
    Nyquist=struct;
    k=struct;
    [Nyquist.x,k.x]=FourierScaledAxis(Number.x,Length.x);
    [Nyquist.y,k.y]=FourierScaledAxis(Number.y,Length.y);
    [Nyquist.z,k.z]=FourierScaledAxis(Number.z,Length.z);
    [k.x,k.y,k.z]=meshgrid(k.x,k.y,k.z);
end
function [Nyquist,k]=FourierScaledAxis(Number,Length)
    d=Length/(Number-1);
    Nyquist=0.5/d;
    N1=(Number-1)/2;
    k=(-N1:N1)*(2*Nyquist/Number);
end
function [Length,Number]=Extract_Coordinates(Grid_3D)
    Length=struct;
    Temp=Grid_3D.x(:);
    Length.x=max(Temp)-min(Temp);
    Temp=Grid_3D.y(:);
    Length.y=max(Temp)-min(Temp);
    Temp=Grid_3D.z(:);
    Length.z=max(Temp)-min(Temp);

    Number=struct;
    [Number.x,Number.y,Number.z]=size(Grid_3D.x);
end

%% Rotations
function Protein=RotateStructureIndex(F,R)
    N=max(size(F));
    [x,y,z]=meshgrid(((1:N)-(N+1)/2)/(N/2));
    Q=R*[x(:),y(:),z(:)]';
    Qx=reshape(Q(1,:),[N N N]);
    Qy=reshape(Q(2,:),[N N N]);
    Qz=reshape(Q(3,:),[N N N]);
    Protein=interp3(x,y,z,F,Qx,Qy,Qz,'linear',0);
end
function Q=Uniform_SO3_Hopf(N_orient)
    n1=round(N_orient^(1/3));
```

```matlab
        N_orient=n1^3;
        N=[n1 n1 n1];

        Psi_=(2*pi)*linspace(0,1,N(1)+1);
        %Theta_=acos(Factor*linspace(1,-1,N(2)+1));
        Theta_=acos(linspace(1,-1,N(2)));
        Phi_=(2*pi)*linspace(0,1,N(3)+1);

        Psi_=Psi_(1:N(1));
        Theta_=Theta_(1:N(2));
        Phi_=Phi_(1:N(3));

        Psi_=Psi_-mean(Psi_);
        Theta_=Theta_-mean(Theta_)+(pi/2);
        Phi_=Phi_-mean(Phi_);

        [Psi,Theta,Phi]=meshgrid(Psi_,Theta_,Phi_);
        Psi=Psi(:);
        Theta=Theta(:);
        Phi=Phi(:);

        Index=1:N_orient;
        Psi=Psi(Index);
        Theta=Theta(Index);
        Phi=Phi(Index);

        Q=[cos(Theta/2).*cos(Psi/2),...
          sin(Theta/2).*sin(Phi+Psi/2),...
          sin(Theta/2).*cos(Phi+Psi/2),...
          cos(Theta/2).*sin(Psi/2)]';
        Q=UnitMagPos(Q);
end
function Q=UnitMagPos(Q)
    Norm=zeros(1,size(Q,2));
    for cntr=1:4
        Norm=Norm+Q(cntr,:).^2;
    end
    Norm=sqrt(Norm);
    Sign=sign(Q(1,:));
    for cntr=1:4
        Q(cntr,:)=Q(cntr,:).*Sign./Norm;
    end
end
function R=Axis2RotMatBatch(Axis)
    N=size(Axis,2);
    R=zeros(9,N);
    for cntr=1:N
        R(:,cntr)=reshape(Axis2RotMat(Axis(:,cntr)),[9 1]);
    end
    if sum(isnan(R(:)))
        disp('Nan in rotation matrix batch')
    end
end
function Q=Uniform_SO3_PDF(N)
    Q=randn(4,N);
    Norm=sqrt(Q(1,:).^2+Q(2,:).^2+Q(3,:).^2+Q(4,:).^2);
    for cntr=1:4
        Q(cntr,:)=Q(cntr,:)./Norm;
    end
end
function R=AllRotMatrices(N)
    Mode=2;
    switch Mode
        case 1  %Random
            Q=Uniform_SO3_PDF(N);
            Axis=Quat2Axis(Q);
```

```matlab
        case 2  %Hopf
            Q=Uniform_SO3_Hopf(N);
            Axis=Quat2Axis(Q);
    end
    R=Axis2RotMatBatch(Axis);
    save Axis.mat Axis  '-v7.3';
end
function R_orth=Polar_Decompose(R)
    [U,~,V]=svd(R);
    R_orth=U*V';
end
function R=Axis2RotUnfold(Axis)
    N_R2=9;
    N_orient=size(Axis,2);
    R=zeros(N_R2,N_orient);
    for cntr=1:N_orient
        Temp=Axis2RotMat(Axis(1:3,cntr));
        R(1:N_R2,cntr)=Temp(:);
    end
end
function Q=RotMat2Quat(R)
    Axis=RotMat2Axis(R);
    Q=Axis2Quat(Axis);
end
function R=Axis2RotMat(Axis)
    Theta=norm(Axis);
    Axis=Axis/Theta;
    a=cos(Theta);
    la=1-cos(Theta);
    b=sin(Theta);
    m=Axis(1);
    n=Axis(2);
    p=Axis(3);
    R=[a+m^2*la, m*n*la-p*b, m*p*la+n*b; ...
        n*m*la+p*b, a+n^2*la, n*p*la-m*b; ...
        p*m*la-n*b, p*n*la+m*b, a+p^2*la];
end
function Q=Axis2Quat(Axis)
    if size(Axis,1) > 3
        Axis=Axis';
    end
    N=size(Axis,2);
    Q=zeros(4,N);
    for cntr=1:N
        Temp=Axis(:,cntr);
        Theta=norm(Temp);
        if Theta
            Q(:,cntr)=[cos(Theta/2);sin(Theta/2)*(Temp/Theta)];
        else
            Q(:,cntr)=[1;0;0;0];
        end
    end
end
function Axis=RotMat2Axis(R)
    x=R(3,2)-R(2,3);
    y=R(1,3)-R(3,1);
    z=R(2,1)-R(1,2);
    r_2sin=norm([x,y,z]);
    if r_2sin
        Theta=atan2(r_2sin,trace(R)-1);
        Axis=(Theta/r_2sin)*[x;y;z];
    elseif R==eye(3)
        Axis=[0;0;0];
    else
        disp('Problem with the rotation matrix')
        R
```

```matlab
        end
    end
    function Axis=Quat2Axis(Q)
        N=size(Q,2);
        Axis=zeros(3,N);
        for cntr=1:N
            Axis(:,cntr)=Quat2AxisSingle(Q(:,cntr));
        end
    end
    function Axis=Quat2AxisSingle(Q)
        Angle=real(2*acos(abs(Q(1))));
        if ~Angle
            Axis=zeros(3,1);
        else
            Axis_norm=Q(2:4)/norm(Q(2:4));
            Axis=Angle*Axis_norm;
        end
    end
```